# Mid-Infrared Drift Scanning Up The SNR Slope

Christopher Packham*[a, b], Amílcar R. Torres-Quijano[a], Sergio Fernandez Acosta[c]

[a]Department of Physics & Astronomy, University of Texas at San Antonio, 1 UTSA Circle, San Antonio, Texas, 78249, USA; [b]National Astronomical Observatory of Japan, 2-21-1 Osawa, Mitaka, Tokyo 181-8588, Japan; [c]GranTeCan, Instituto de Astrofisica de Canarias, C/ Vía Láctea, s/n E-38205 La Laguna (Tenerife), Spain

## ABSTRACT

Mid-infrared (MIR) observations are typically accomplished from the ground through oscillating the secondary mirror a few times a second. This chopping serves to remove the fast time variable components of (a) sky variation and (b) array background. However, there is a significant price to pay for this, including reduced on-object photon collection time, stringent demands on the secondary mirror, nodding the telescope to remove the radiative offset imprinted by the chopping, and an often-fixed chop-frequency regardless of the sky conditions in the actual observations. Worse, in the era of 30m telescopes it is wholly impracticable to chop the secondary mirror. If the array is stable enough, drift scanning holds the promise to remove the necessity of chopping. In this paper we report our experiments using the CanariCam MIR instrument on the 10.4m GranTeCan and the implications to future instruments and experiments.

## 1. INTRODUCTION

In this paper we discuss the history of the CanariCam MIR instrument on the GranTecan, chopping and nodding, and an overview of drift scanning (§1). In §2 we discuss our experimental design and observing log, and in §3 we detail our data reduction approach and steps, including the results. In §4 we comment on our conclusions, and §5 briefly presents our discussion and future work plans.

### 1.1 History of CanariCam

CanariCam (Telesco et al., 2003) is the facility multi-mode MIR (8-25 μm) camera on the 10.4-m Gran Telescopio CANARIAS (GTC) on La Palma, Spain. It was designed and built by the University of Florida (PI: Charles Telesco) and affords the community imaging, spectroscopic and unique polarimetric MIR capabilities at, or near, the diffraction limit of the telescope. Since 2012, it had been operating in queue mode at one of the Nasmyth focal stations, until it was temporarily decommissioned in 2016. Following an upgrade project (Fernandez-Acosta, 2020) started in 2018, it was reinstalled and recommissioned in a folded-Cassegrain focal station in late 2019, retrofitted with a more powerful and reliable 4 Kelvin cryocooler and new electronics and interfaces.

Thanks to its versatile instrumental modes, the large collecting power of the GTC and its high spatial resolution, CanariCam served a broad range of science cases in fields spanning Active Galactic Nuclei (i.e. Alonso-Herrero et al 2011), Supernovae (i.e. Telesco et al 2015), the Galactic Center (Roche et al 2018), Protoplanetary Disks (i.e. Li et al 2016) and Substellar Objects among others, and its polarimetric modes (Packham, Hough, & Telesco 2005) were unique and very successful, enabling the study of magnetic fields through the absorption or emission of aligned dust grains. It has produced 44 refereed papers, with 534 citations and an H-Index of 13, from a total of 876 hours of telescope time. After deducting the hours delivered during the last year, in accordance with the GTC metrics to account for publication lag, it corresponds to a good productivity ratio of 18 hours per paper, in comparison to the average of 23 h/paper from GTC open-time proposals (Cabrera-Lavers, 2020). It must be noted that MIR observations are inherently less efficient than those in the visible wavelengths, due to the overheads of the chop-nod technique; and that the demand from the GTC community in the MIR represents only a small fraction of the telescope overall production, more biased towards optical wavelengths.

Due to the very demanding instrumentation program at GTC, and the need to prepare for, and host, the next generation instruments, CanariCam has been decommissioned from the telescope in January 2021.

*chris.packham@utsa.edu

## 1.2 The Chop/Nod Technique

As noted in §1.1, MIR observations typically require the secondary mirror of the telescope to be oscillated at a frequency of a few times a second to subtract the background. The frequency of the chopping (for the current generation of MIR arrays) is typically set by the array so-called (very fast changing) "1/f" noise, but also removes the time-variable sky background (fast changing) and telescope thermal emission (slow changing). In the N band (~7.7-13μm) at a good site and under good conditions an approximate estimate for the characteristic sky emission stability is ~0.2 to 20 seconds (dependent on the wavelength and conditions) and the telescope thermal emission stability is >~20 seconds. Figure 1 shows the relative power spectrum of the MIR sky in the N band as ESO/La Silla using TIMMI2.

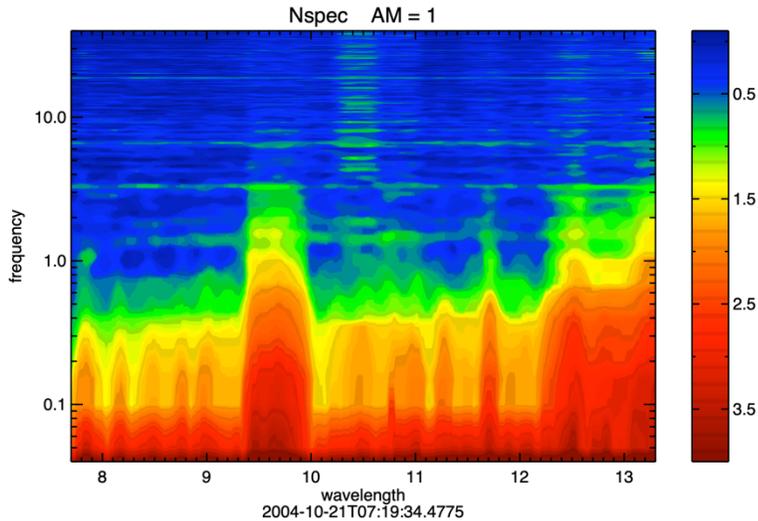

Figure 1.: Relative power spectrum of the sky, showing the variation of the sky through the N band (Pantin).

However, the chopping of the secondary induces several deleterious effects to the data and data collection efficiency including:

1.  Fast guiding is often available in only one beam of the chopped image, sometimes rendering the unguided beam unusable due to the degraded point spread function (PSF). Thus, the photon collection time is reduced by this single effect by 50%.

2.  Reduced on-source photon collection time, as the mirror is physically moving and settling to a stable position for a fraction of the elapsed (clock) time.

3.  As the beam propagates through the telescope through two slightly different positions, the telescope emission is not precisely the same, leaving a so-called radiative offset. This can be countered by nodding the telescope in an equal and opposite amount of the chop. The cadence of this nodding is typically a few 10's of seconds, dependent on the rotation of the pupil and the centrosymmetry of the pupil, which can also imprint a radiative offset on the final data product. The time required for nodding and subsequent settling further reduces the photon collection time.

4.  As the secondary mirror is partially de-collimated by the act of chopping, optical aberrations are induced in the chopped beam, gaining in severity as a function of chop distance. This is shown in Figure 2 for the GTC/CC system.

5.  It is possible that the off-source (chopped) beam can contain other sources and potentially extended sources emission, therefore resulting in incorrect sky subtraction.

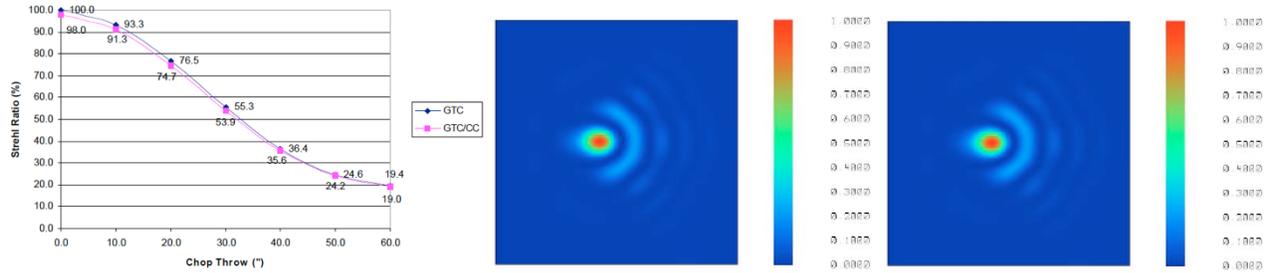

Figure 2.: Effect of the chop throw on CanariCam Strehl ratio (left). Image delivered by the GTC (middle) and CanariCam/GTC (right) for a chop throw of 30", shown in log-scale.

These effects place tight constraints on MIR astronomy, and the reduction in on-source photon collection time can be significant. In the case of the GTC/CC, and where the chopped beam is excluded, the observing efficiency (defined as photon collection time/clock time) is ~32%, typical for MIR observations. If the chopped beam (the 'negative' beam) can be included to the guided beam (the 'positive' beam), the observing efficiency rises to ~64%. Finally, the problems above do not pay any regard to the demanding engineering requirements that a rapidly oscillating mirror imposes on the telescope. Indeed, for the 30m class telescopes, the secondary mirror will be too large to chop (i.e. in the case of the TMT, M2 is 3.1m is diameter) and hence other methodologies for removal of the array/sky noise are required. Instead, promising work proceeds for chopping/nodding a mirror located at a pupil plane interior to the instrument (i.e. Honda et al. 2020), principal component analysis (PCA, Hunziker et al., 2017), and so called drift scanning (i.e. Ohsawa et al., 2018).

### 1.3 Drift Scanning & Array Issues

Drift scanning makes use of the MIR array capabilities to be rapidly readout whilst the telescope is either held at a single position with respect to the mount (i.e. not tracking the projection of the moving sky) or being driven at a user defined speed across the sky field (i.e. tracking plus/minus an increment in RA or Dec). Thus, each image contains a short exposure (measured in units of ms) which can be differenced from a later obtained frame and then shifted and co-added in subsequent data reduction. The secondary remains fixed, so the observing efficiency rises to ~100%, and the sky is removed by subtraction of earlier or subsequent frames. All optical aberrations are completely eliminated as compared to chopping, as are guiding issues in the chopped beam, but unless the drift scan speed is extremely fast (greater than a few seeing discs per second), the noise subtraction will be inferior (in the absence of more advanced data reduction techniques, such as PCA). However, if the drift scanning is too fast, the image of the object will be elongated along the scanning direction hence an optimal balance maybe be made.

CanariCam uses a CRC774 Raytheon array as the MIR sensor. Consisting of 320x240 pixels, it was state of the art at the time of CanariCam's design. However, it is well known than this array suffers from several array artifacts due to the IR sensitive materials and multiplexer used (Sako et al 2003). CanariCam utilizes a readout technique to mitigate those effects, but residuals are still present. In standard chop/nod mode, the primary effect is a "cross-talk" in each of the 16 channels of 20 pixels width, where a bright object in one channel causes a negative image in each of the other 15 channels, but at a fraction of the object flux. Thus, a negative image in the same x/y position of the object is produced. The inverse is true for the chopped beam, where the negative beam creates a positive cross-talk effect. When drift scanning, this effect is also present. As the image is the differenced frame from closely separated temporal images, a pair of differenced images contains a positive and negative image of the source that are spatially close. In this paper, we compare the results of drift scanning of a photometric standard with those made in standard chop/nod mode, temporally separated by at most a few hours.

## 2. EXPERIMENT DESIGN & DATA COLLECTION

### 2.1 Observational Setup

While the drift-scanning is not an observing mode offered by GTC, standard non-sidereal tracking and guiding functionality is flexible enough to carry out a drift-scanning-like observation with minimal effort. By creating an artificial ephemeris file for our sidereal object, specifying virtual positions in astrometric right ascension and declination coordinates, and velocities at tabulated time intervals, we can drift it at the desired rate following an arbitrary path across the FOV. The main advantage of this highly flexible approach is that these technical observations did not require any modification to the GTC Control System (Filgueira & Rodriguez, 1998), and still allowed to characterize the potential of this mode for future developments and experiments.

The operational drawbacks identified were all related to managing the synchronization between the telescope non-sidereal tracking and the acquisition process on the instrument field of view, since, ideally, one would like to begin the drift from a specific known position on the detector when the detector starts exposing. Also, the ephemeris file must be generated ad-hoc for the target and for a given time slot, because the drift is to be performed on sky RA and DEC coordinates, and we created a Python script to automate this process. For a future development, the drift pattern could be programmed at the GCS-level using the focal plane coordinate system (fixed with respect to the instrument), which would make it more generally applicable to any target on-sky and easier to synchronize with the instrument acquisition.

## 2.2 Observation Description & Data Log

To compare the signal to noise ratio (SNR) obtained in imaging mode, we made observations of the same standard star in (1) standard chop/nod mode and (2) drift scanning mode, temporally separated by only a few minutes. Assuming background stability, this enables comparison of those data sets and hence characterizes the difference in SNR for both (a) on-source exposure time and (b) clock elapsed time. For this experiment, the background is the addition of sky, telescope, window, array, and all other sources.

The source chosen was of sufficient brightness to obtain excellent SNR in both a short chop/nod exposure, as well as individual images within the drift scan stack. When obtaining the drift scan, the telescope was driven in non-sidereal tracking at 3 different speeds. In all cases, the star was moved to describe a rectangular shape across the 320x240 array, because it would allow us:

- to characterize performance for the two drift directions independently (whether drifting along the detector channels or across them have any impact in the final instrument sensitivity due to the detector artifacts), and

- to minimize the synchronization requirements between the telescope drift and the instrument integration because the target will always be in the FOV and following the same path for a long period of time.

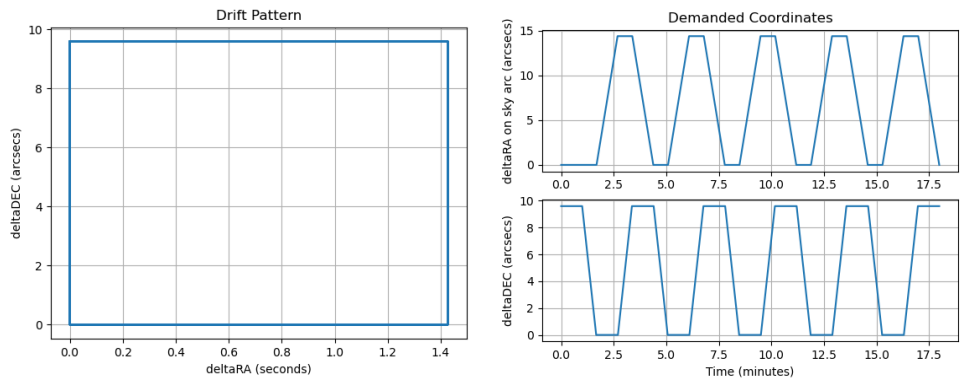

Figure 3. Representation of the ephemeris file and demanded astrometric coordinates for one of our observations.

We drove at 3 different speeds to empirically evaluate which is the optimal speed from SNR and full width at half maximum (FWHM) metrics for our given environmental conditions and instrumental setup. The speed of drifting was set to be a multiple of the seeing disk, and through this experiment we try to balance the degradation of the FWHM vs. the improvement of obtaining better sampled background subtraction.

Due to limitations in the data acquisition system, the detector controller cannot transfer all individual frames, with typical integration times of 25 ms, but it must accumulate a certain number of them in its internal buffer, and finally transfer and save an accumulated image, called "saveset", at a maximum rate of approximately 5 Hz (savetime=0.207s). This constrains the maximum drift rate at which the PSF will not be elongated in each saveset, and we chose the maximum drift as 3 pixels/s, which will only degrade the PSF a fraction of a pixel. This is the main limiting factor of CanariCam for the drift-scanning mode because this will ultimately limit the frequency at which we can subtract the highly variable background.

Assuming median conditions at the Observatorio del Roque de Los Muchachos, we expected an image quality with a FWHM around 0.32 arcsec, which corresponds to 4 pixels on the CanariCam array, with a plate scale of 0.08"/pix. With a 3 pix/s drift, the PSF will move one seeing disk every 1.33 seconds, while with a standard chopping configuration of 2Hz, the background can be subtracted every 0.25 seconds, better sampling its variations along the observation.

A special engineering "burst" mode could be used with CanariCam in late 2019, enabling that all the individual frames were transferred and stored, leading to a savetime of only 0.0253 s. In this configuration we could have used a much higher drift rate to better sample and subtract the background while still preserving the PSF in the individual frames, but the observing conditions were too poor (seeing ranging from 2" to 4" in the visible) to exploit this capability at all, and this run was mainly devoted to test and optimize the acquisition method. However, the MIR conditions were good and stable, and thus the background is assumed to be similar. In this paper we describe only the 'traditional', slow readout method of CanariCam, and in a subsequent paper we will analyze the 'burst-mode' observations. The precipitable water vapor (PWV) is noted in the comments section of the observing log below.

Table 1. Log of observations

| Date | UTC at start | Elapsed time (s) | On-Source Time (min) | Savetime (s) | Object | Filter | Observing Mode / Setup | Comments |
|------|------|------|------|------|------|------|------|------|
| 28/07/2020 | 01:04:00 | 619.55 | 3.31 | 5.79 | HD213310 | Si2-8.7 | Chop-Nod 5.7 arcsec North (up) Chop Freq. = 2.07Hz | Reference data before drift scan experiment to measure sensitivity. Good image quality. PWV=6.5 mm |
| 28/07/2020 | 01:31:03 | 600.06 | 10.001 | 0.207 | HD213310 | Si2-8.7 | Stare. Drift rate = 0.24 arcsec/s | Drift scanning #1 at fast rate of ~3 pix/s. PWV between 6.3 and 5.7 mm |
| 28/07/2020 | 01:51:13 | 600.06 | 10.001 | 0.207 | HD213310 | Si2-8.7 | Stare Drift rate = 0.08 arcsec/s | Drift scanning #2 at medium rate of ~1 pix/s. PWV=6.2 mm |
| 28/07/2020 | 02:02:18 | 600.06 | 10.001 | 0.207 | HD213310 | Si2-8.7 | Stare Drift rate = 0.08 arcsec/s | Repeat. PWV=6.2 mm |
| 28/07/2020 | 02:17:03 | 600.06 | 10.001 | 0.207 | HD213310 | Si2-8.7 | Stare Drift rate = 0.04 arcsec/s | Drift scanning #3 at slow rate of ~0.5 pix/s. PWV=5.8 mm |
| 28/07/2020 | 02:27:14 | 600.06 | 10.001 | 0.207 | HD213310 | Si2-8.7 | Stare Drift rate = 0.04 arcsec/s | Repeat. PWV between 5.8 and 6.0 mm |
| 28/07/2020 | 02:42:32 | 401.49 | 2.137 | 1.45 | HD213310 | Si2-8.7 | Chop-Nod 5.7 arcsec North (up) Chop Freq. = 2.07Hz | Reference data after drift scanning. Good image quality. PWV between 6.0 and 6.3 mm |
| **First Drift Scanning Experiments under very poor seeing conditions (2"-4") using Burst Mode (savetime=frmtime)** | | | | | | | | |
| 08/11/2019 | 03:03:09 | 179.9 | 2.998 | 0.0363 | HD36167 | Si2-8.7 | Stare in burst mode. Drift rate = 0.15 arcsec/s | Drift Scanning in burst mode. PWV=2.9 mm. Slow Guiding |
| 08/11/2019 | 03:08:53 | 191.796 | 3.1966 | 0.0363 | HD36167 | Si2-8.7 | Stare in burst mode. Drift rate = 0.15 arcsec/s | PWV=2.9-3.0 mm. Slow Guiding |
| 08/11/2019 | 03:14:37 | 191.951 | 3.199 | 0.0253 | HD36167 | Si2-8.7 | Stare in burst mode. Drift rate = 0.15 arcsec/s | PWV=3.0 mm. Fast Guiding |

# 3. DATA REDUCTION

## 3.1 Data Reduction Approach & Results

We divide the background flux (B) into three sources, the sky (S), telescope thermal emission (T), the instrument background (I, including entrance window), and the array bias (A). The object (O) was drift scanned at three different rates across the array (see section 2). Thus, a single frame consists of the background at frame number n ($B_n = S_n + T_n + I_n + A_n$), where n is the frame at number (n) = 1-2,901 for our longest drift scan. We can then form a difference image where we account for $B_n$ by subtracting a frame $F_{n+\delta n}$ where $\delta n$ is a frame taken subsequently. T and I tend to zero significant emission change during the course a single observation, and hence are readily removed through. S and A can vary at short timescales, and this implies that $\delta n$ should tend to as lower integer as possible to sample that background. However, to maintain photometric integrity the $\delta n$ must be large enough to ensure the point spread function (PSF) of $O_n$ and $O_{\delta n}$ are sufficiently distant so as to cause negligible photometric effects on each other (i.e. the positive and negative differenced images of the source must be sufficiently distant). This drives $\delta n$ to a value that is dependent on the drift scan speed and the delivered PSF of the instrument/telescope combination, and is much greater than that which would be desired to sample B. In our observations, the conditions were excellent and the delivered PSF was close to the diffraction limit, hence the limits on the elongation of the PSF were close to the theoretically most stringent. If the drift scan speed is too high, the PSF will not be sampled by $F_n$ and will be elongated in the direction of drift, hence we obtained drift scans at three different speeds to ensure the PSF was sampled sufficiently to eliminate the elongation but also provide sufficient distance between the positive and negative images and also sample S and T at sufficient temporal resolution.

We proceeded to reduce the data in a staged approach for each of the three drift scan files. A background corrected image was obtained by the difference of $F_n - F_{n+\delta n}$ where $\delta n$ is varied to obtain the parameters as described below:

1. Determine the residual noise per pixel for a range of $\delta n$. $\delta n$ was set to 1, and then all frames (from 1-2,901) were differenced and coadded and the noise statistics measured in that final image. $\delta n$ was incremented by an integer and the process repeated. The noise per pixel vs. $\delta n$ is shown in Figure 6. The noise was measured in the frame at a sufficient distance from the source (Figure 4) to ensure no contamination from the source object nor negative images. As noted in the introduction, the positive and negative images are taken both temporally and spatially close to each other. With this experiment, we drift scanned in a rectangle, thus the images trace a +x, +y, -x, then -y telescope motion, thereby providing a central positive image (as we used the centroid of the positive image as our origin when co-adding frames) and four negative images representing the four drift scan motions. This is shown in Figure 4.

2. Determination of the aperture size. We computed a differenced frame of $\delta n$ where the distance between the positive and negative images was 15 times the measured PSF to ensure that a photometric aperture could be placed over the positive image without effect from the cross-talk. For this criteria to be met, $\delta n=45$ and we differenced and then coadded all frames. To remove the effects from the negative images and cross-talk, we applied a threshold to the image, where any pixels ≤ (mean – (3 times the standard deviation) was set to zero. Next, a circular aperture was placed over the image with increasing aperture radius, and the curve of growth of the constrained flux was plotted (Figure 7). At a radius of 40 pixels this plot shows that we can safely assume that all flux is contained within the aperture. We set the aperture size where 95% of the total flux was contained within the aperture, which was determined to be 10 pixels.

3. Determine the optimal value of $\delta n$ to ensure photometric integrity, where the negative image has negligible impact on the photometric values. $\delta n$ was set to 1 and then all frames (from 1-2,901) were differenced and a circular aperture of 10 pixels was placed over the positive image and the flux in that final image was measured. $\delta n$ was incremented and the process repeated. The counts contained within the aperture vs. frame separation is shown in Figure 8. Where $\delta n >= 40$, the increase in flux is <0.1%, so we consider that a $\delta n=45$ contains all the flux and is essentially unaffected by the negative image. We then found $\delta n$ where >=95% of that flux is constrained in the aperture, which is $\delta n=26$.

4. A final coadded image where $\delta n=26$ was constructed. We extracted an 80x80 pixel sub-frame, where the center of the image was the centroid of the positive frame. Sub-pixel shifting/rebinning was not performed to avoid image quality degradation and flux redistribution. Only where >50% of the pixel was contained within the circular aperture was the pixel's counts included in the total. The noise was estimated from stage 1 above, and thus the SNR could be computed, using the equation below. Finally, the full width at half maximum (FWHM) was measured by fitting a Moffat profile to the final image to provide an estimate of the delivered PSF.

SNR = source_counts / (noise_per_pixel x (number_of_pixels_in_aperture)$^{0.5}$)

5. <u>Chop/nod data reduction</u>. The standard chop/nod image is shown in Figure 5 below, reduced in the typical reduction methodology.

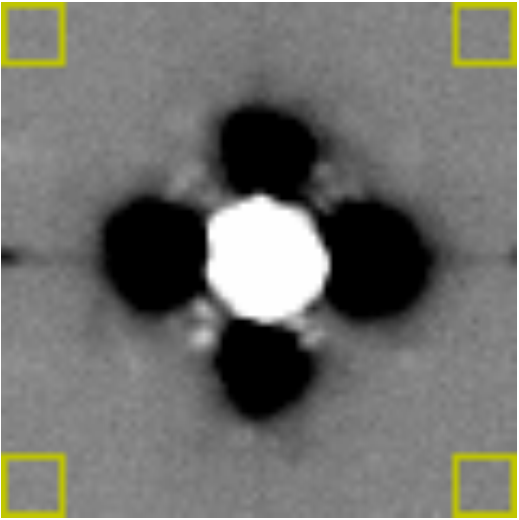

Figure 4. Drift Scan coadded image at δn=26. The yellow boxes located at the corners of the image represent the regions utilized to calculate the noise per pixel. These boxes are 10x10 pixels in size. The boxes were placed at this location to avoid the negative count value contributions from the cross-talk and negative beams.

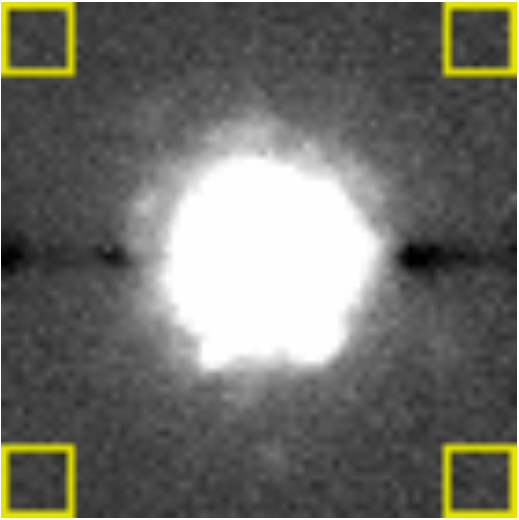

Figure 5. Chop/nod reduced image. The yellow boxes located at the corners of the image represent the regions utilized to calculate the noise per pixel. These boxes are 10x10 pixels in size. The boxes were placed at these locations to avoid the negative count value contributions from the cross-talk and negative beams.

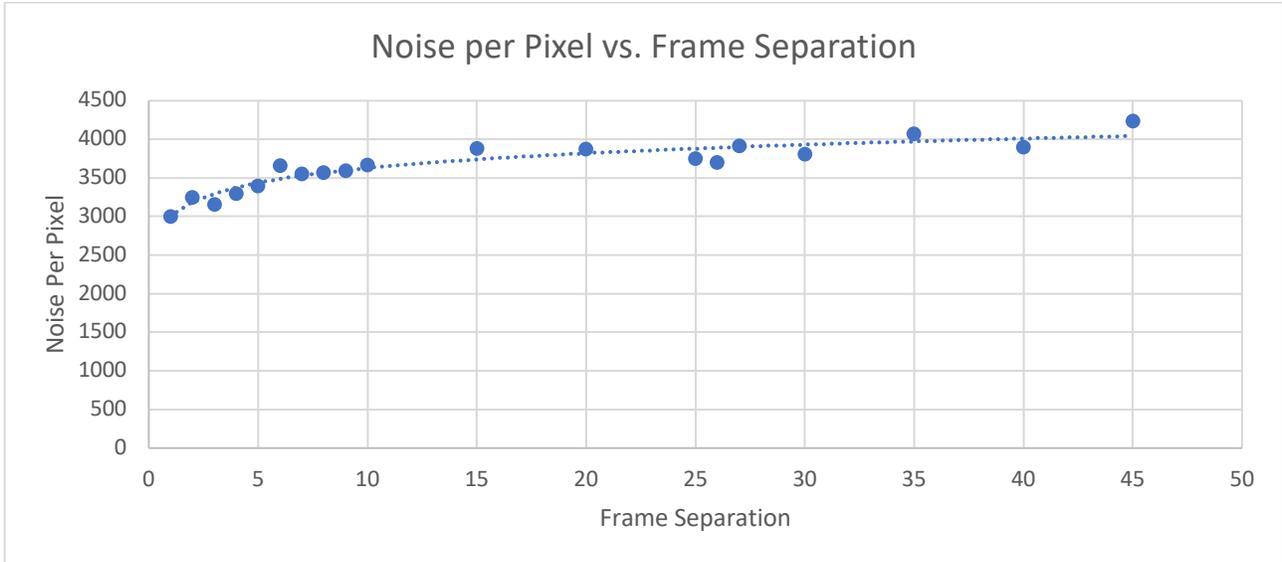

Figure 6. Evolution of the noise per pixel of the coadded drift scan images as a function of frame separation. The noise per pixel stabilizes at δn=~15.

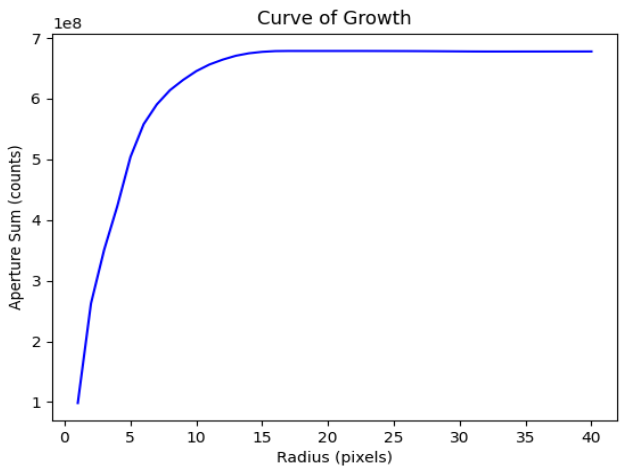

Figure 7. Photometric curve of growth of the total flux at different aperture radii for the coadded drift scan image file at δn=45.

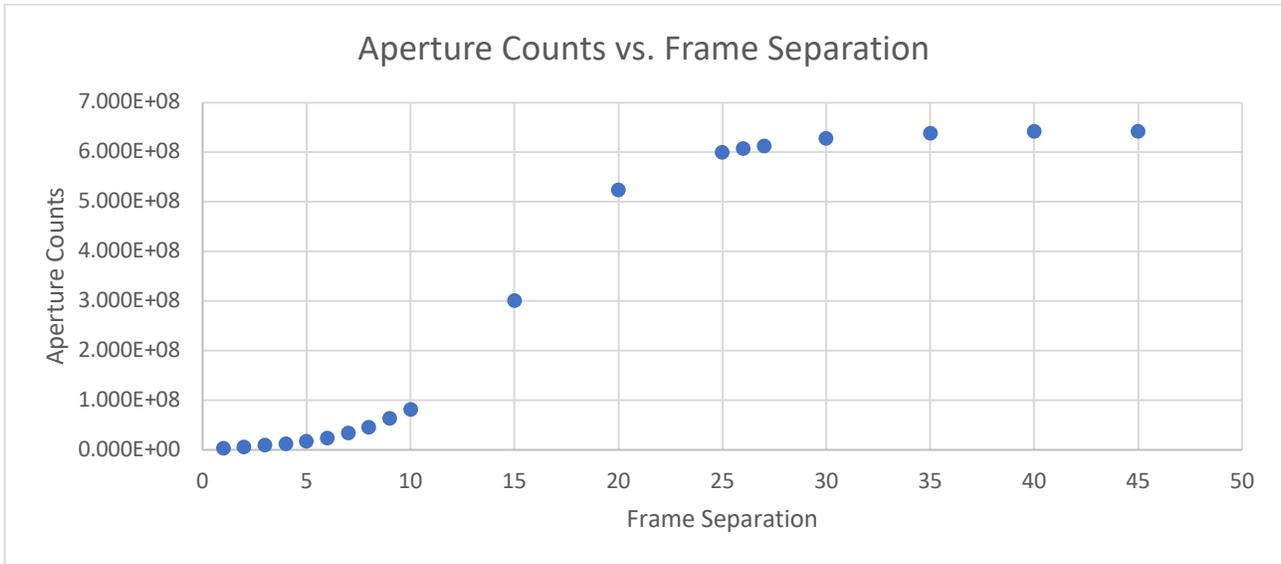

Figure 8. Aperture counts as a function of frame separation. The total flux starts stabilizes at ~δn=~30.

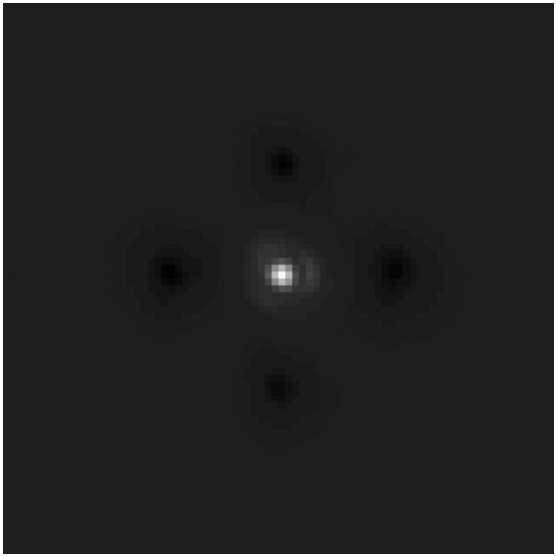

Figure 9. Drift-Scan coadded image for δn=26 which corresponds to the optimal frame separation determined in step 4.

It is clear from the figures above that to ensure photometric integrity, δn=26 is required using our criteria, but this does imply an inferior background noise than if δn<~10. A faster drift scan speed would provide a greater beam separation to ensure photometric integrity and obtain a better background noise value. The measured signal, noise, and hence SNR is shown below:

Table 2. Reduced drift scan image parameters.

| Method | Clock Time (s) | On-Source Time (s) | Noise per pixel (counts) | No. of Pixels | Total Signal (counts) | SNR | FWHM (pixels, ") |
|---|---|---|---|---|---|---|---|
| Drift scan | 600.06 | 600.06 | $3.700 \times 10^3$ | 305 | $6.072 \times 10^8$ | $9.396 \times 10^3$ | 2.79, 0.223 |

As noted in section 2, the same object was observed in chop/nod mode and was reduced in (a) the standard chop/nod reduction method. This leads to a central positive image and two negative images at the location of the chop throw, in this case 15". We also reduced the same dataset performing (b) a shift and add for each data saveset and also (c) inverting the negative image and adding to the positive beam. We measured the noise of the image, ensuring that the effects from cross-talk and the positive/negative images were excluded, and then measured the signal in the same sized aperture that we used for the drift scan data. Standard methodology for MIR observations is to use the chop/nod positive (central) image only (a), but for this work we elected to reduce the data in methods (b, c) to enable a more appropriate comparison to the drift scan data reduction; the most appropriate comparison is to method (c). We tabulate the different methodological parameters and results below.

Table 3. Reduced chop/nod image parameters.

| Method | Clock Time (s) | On-Source Time (s) | Noise per pixel (counts) | No. of Pixels | Total Signal (counts) | SNR | FWHM (pixels, ") |
|---|---|---|---|---|---|---|---|
| Chop/nod | 619.55 | 198.6 | $1.650 \times 10^3$ | 305 | $2.097 \times 10^8$ | $7.277 \times 10^3$ | 3.29, 0.263 |
| Chop/nod with shift & add | 619.55 | 198.6 | $1.649 \times 10^3$ | 305 | $2.102 \times 10^8$ | $7.300 \times 10^3$ | 3.23, 0.258 |
| Chop/nod, shift & add, -ve and +ve beams combined | 619.55 | 397.2 | $2.838 \times 10^3$ | 305 | $4.185 \times 10^8$ | $8.444 \times 10^3$ | 3.30, 0.263 |

Finally, we can compare the drift scan SNR to those of the three chop/nod methods, as summarized below.

Table 4. SNR comparison.

| Method | Clock Time (s) | SNR | SNR Ratio Compared to Drift Scan |
|---|---|---|---|
| Chop/nod | 619.55 | $7.277 \times 10^3$ | 1.291 |
| Chop/nod with shift & add | 619.55 | $7.300 \times 10^3$ | 1.287 |
| Chop/nod, shift & add, -ve and +ve beams combined | 619.55 | $8.444 \times 10^3$ | 1.113 |
| Drift scan | 600.06 | $9.396 \times 10^3$ | - |

We can compare the SNR between the data reduction methods for the chop/nod data to the drift scan data. The typical manner of data reduction used in science observations is that of the chop/nod data reduction, but we include the chop/nod data that has been shifted and added and the negative beams combined to the positive beam to show a similar on-source exposure time. However, arguably the most important metric is comparing the elapsed time, as that is the allocated time on the telescope. We can apply the proportionality of the SNR to (exposure time)$^{0.5}$ to estimate what a simple scaling would suggest. Thus, we can estimate the signal should increase by:

a. Chop/nod: $(600.06/198.6)^{0.5} = 1.73$

b. Chop/nod with positive/negative beams combined: $(600.06/397.2)^{0.5} = 1.23$

The noise, however, will certainly increase as we are sampling the background at a far slower cadence. Thus, the values above are the best case for how much the SNR could be increased.

## 4. CONCLUSIONS

The data and results described within this paper shows that indeed the GTC/CanariCam combination can be used in drift scan mode to obtain science quality data. The SNR and FWHM of the data show an improvement over that of standard chop/nod obtained data (Table 4 and Figure 10). More specifically, the obtained signal is significantly improved in drift scanning, but the noise is worse. The inferior noise performance is likely a limitation of the array/electronics system which does not have the capability to read the data out in very fast frames, such as "burst model" (i.e. the savesets are too long in time to adequately sample the noise) as would be required to sample the array noise (especially the 1/f noise) adequately. We plan to analyze the data further to investigate if PCA can reduce the noise and hence further increase the final SNR. We also hope that this analysis will be able to probe if the limiting factor is the electronics or array which may be able to help inform future designs. Further, as shown in Table 1, we have obtained data in three drift scan speeds which we will fully analyze using the same methodology used in this paper and with PCA to compare and contrast the effects of drift scan speed on SNR and FWHM.

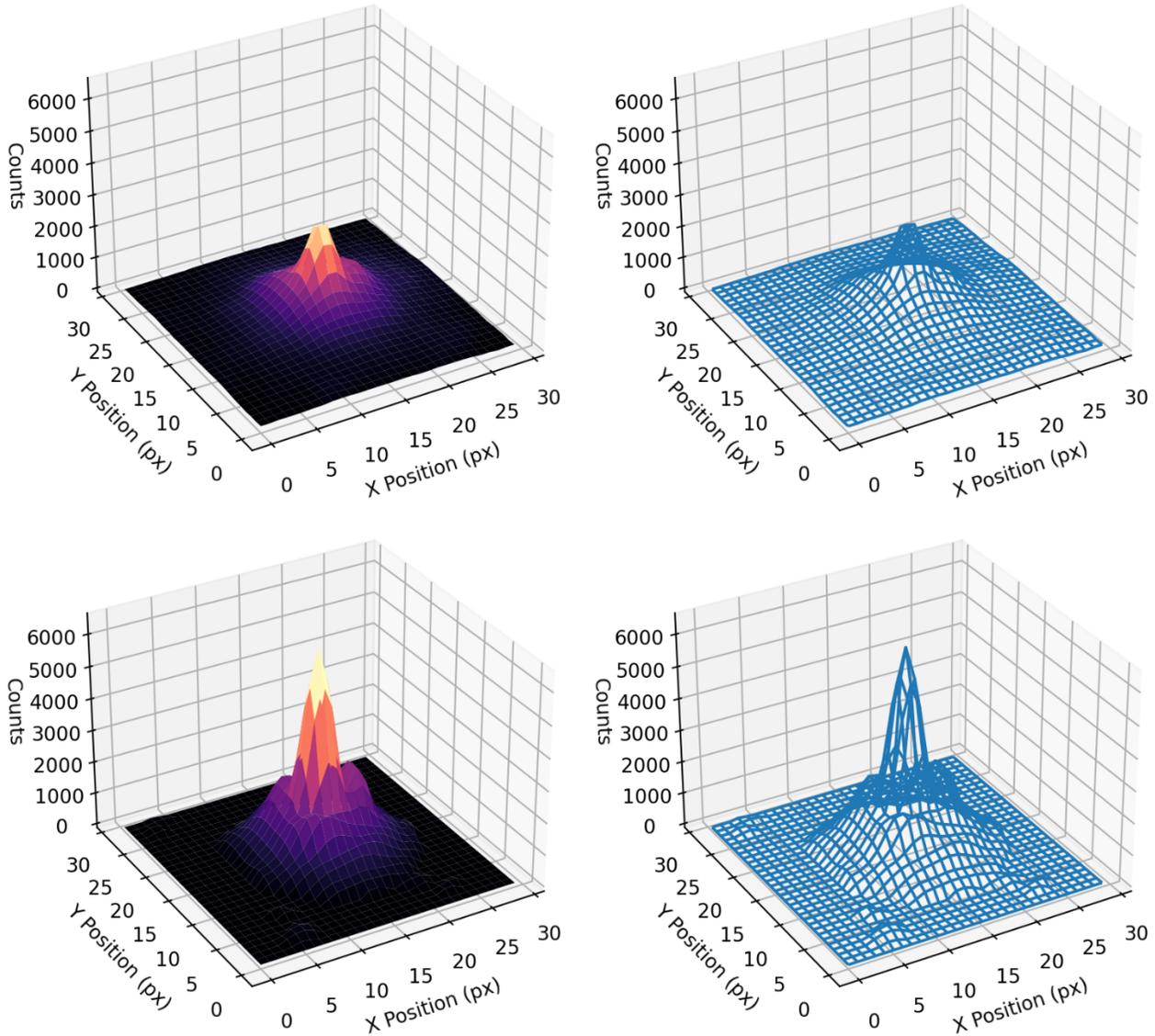

Figure 10. Final chop/nod image (top) and drift scan (bottom) coadded image in color and wire frame representations showing the improved signal and FWHM in the drift scan data. Note the clock-times are very similar and hence are appropriate comparisons.

# 5. DISCUSSION

The GTC/CanariCam system was not designed for drift scanning, but these results demonstrate significant promise in this technique, and subsequent analysis could help inform designs of future instruments, including those envisioned for the 30m class of telescopes. At least some of these instruments plan to use the GeoSnap Teledyne array which has lower noise than the previous generation of arrays (Bowens et al., 2020), and thus may give superior drift scan results. We will continue to work on our unpublished data and perform PCA analysis which will be published in a journal article in due course.

# 6. ACKNOWLEDGEMENTS

Based on observations made with the Gran Telescopio Canarias (GTC), installed at the Spanish Observatorio del Roque de los Muchachos of the Instituto de Astrofísica de Canarias, in the island of La Palma. The upgrade of CanariCam was co-financed by the European Regional Development Fund (ERDF), within the framework of the "Programa Operativo de Crecimiento Inteligente 2014-2020", project "Mejora de la ICTS Gran Telescopio CANARIAS (2016-2020)". The authors wish to acknowledge the help of the entire GTC staff (science and engineering) as well as those key people in the CanariCam project, originated at the University of Florida. ARTQ wishes to acknowledge John Kucewicz and Sharon Smith-Kucewicz for their generous financial assistance received in support of his graduate studies. ARTQ also acknowledges support from the Department of Physics & Astronomy at UTSA. CP expresses his deep thanks to SPIE for accepting his late abstract and manuscript submissions.